\documentclass[aps,prl,twocolumn]{revtex4}

\newcommand{\be}{\begin{equation}}
\newcommand{\ee}{\end{equation}}

\begin{document}

 \begin{flushleft}
Brown-HEP-1651
\end{flushleft}

\large

\title{Where Have All the Goldstone Bosons Gone?}

\author{G. S. Guralnik\footnote {mail to: gerry@het.brown.edu}}

\affiliation{Department of Physics\\
Brown University\\
Providence, R.I.  02912}


\author{C. R. Hagen\footnote{mail to: hagen@pas.rochester.edu}}

\affiliation{Department of Physics and Astronomy\\
University of Rochester \\
Rochester, N. Y. 14627}

\begin{abstract}
According to a commonly held view of spontaneously broken symmetry in gauge theories, troublesome Nambu-Goldstone bosons are effectively eliminated by turning into longitudinal modes of a massive vector meson.  This note shows that this is not in fact a consistent view of the role of Nambu-Goldstone bosons in such theories.  These particles necessarily appear as gauge excitations whenever they are formulated in a manifestly covariant gauge.  The radiation gauge provides therefore the dual advantage of circumventing the Goldstone theorem and making evident the disappearance of these particles from the physical spectrum.      
\end{abstract}



\maketitle

\vskip 1cm

The impact of spontaneous symmetry breaking on current understanding of the structure of particle physics has been nothing less than remarkable.  In three papers Englert and Brout [1], Higgs [2], and Guralnik, Hagen, and Kibble [3] showed that spontaneously broken symmetry leads to the massification of gauge fields.  The troublesome issue of the massless Nambu-Goldstone (NG) boson,  long known to be a severe constraint on such theories, was dealt with in Ref.[3] (GHK) by using the radiation gauge formulation of the theory.  This approach rendered the Goldstone theorem inapplicable and led to an explicit proof of the absence of massless modes.  A covariant gauge was employed in Ref.[1] and the issue of NG bosons was not considered.  Higgs \cite{Higgs} did not specify a gauge, but would have encountered NG bosons had a covariant one been used.  However, conventional wisdom seems collectively to have concluded that the NG boson issue simply disappeared in the context of  Refs.[1] and [2].

According to this view the NG boson issue has conveniently vanished by its being transformed into the longitudinal mode of the massive vector meson and thereby becoming no longer an object of interest.  Perhaps the earliest expression of this view is that of  Anderson [4] who stated somewhat prior to the above referenced  work on symmetry breaking that ``...while the boson which appears as a result of the theorem of Goldstone and has zero unrenormalized mass is converted into a finite-mass plasmon by interaction with the appropriate gauge field, which is the electromagnetic field." Fifty years later the Nobel Committee [5] stated ``The Goldstone theorem holds in the sense that that Nambu-Goldstone mode is there but it gets absorbed into the third component of a massive vector field."  This assertion seeks to lay to rest the issue of the NG boson nemesis by saying that just as in GHK which accomplishes this by an explicit display of the entire spectrum, so also  does the new manifestation as a longitudinal mode eliminate the NG boson as an ongoing concern of the theory.  It is shown in what follows that that is not a valid view and that a massless gauge particle necessarily remains in the theory. A brief review of the simple model which was considered in Refs. [1-3] is given below.

  The fields that describe the system are (in the notation of GHK) the gauge fields $F^{\mu\nu}$ and $A^{\mu}$ together with Hermitian spin zero fields described by $\varphi_i$ and $\varphi^{\mu}_i$ ($i=1,2$).  After the imposition of the broken symmetry condition, 
the equations of motion reduce to 
$$F^{\mu\nu}=\partial^{\mu}A^{\nu} -\partial^{\nu} A^{\mu}$$
$$\partial_{\nu}F^{\mu\nu}=\varphi^{\mu}\eta$$ 
$$\varphi^{\mu}=-\partial^{\mu}\varphi-\eta A^{\mu}$$
$$\partial_{\mu}\varphi^{\mu}=0$$
where 

  $$ \eta=         \left( \begin{array}{c}
                       \eta_1  \\ \eta_2
\end{array}              \right  ).$$
Taking $\eta_2=0$, with no loss of generality,  in the radiation gauge the equations for $\varphi_1$ and $A_k^T$ take the form 
$$(-\partial^2 +\eta_1^2)\varphi_1=0$$

and 

$$(-\partial^2 +\eta_1^2)A_k^T=0$$
where the superscript $T$ denotes the transverse part.  The two degrees of freedom of $A_k^T$ together with $\varphi_1$ comprise the three components of a massive vector field. 
The $\varphi_2$ field is readily given a mass through a suitable ``Mexican hat" potential or by iteration to higher order.  It may well be embodied by the recent discovery at the LHC of a particle of mass 125 GeV, but is not relevant to the task of identifying the role of the NG boson.  

The Green's function of the $\varphi_1$ field is determined using the equation for the divergence of $F^{\mu\nu}$ to obtain 
$$-\nabla^2 A^0 =\eta \varphi^0_1$$
and thus that 
$$\varphi^0_1=(1-\frac{\eta_1 ^2}{\nabla^2})^{-1}(-\partial^0 \varphi_1).$$
The equal time commutator
$$[\varphi^0_1({\bf x},\varphi_1(0)]=-i\delta({\bf x})$$
then implies that the propagator
$$G(x)\equiv i\langle 0|(\varphi_1 (x) \varphi_1(0))_+|0\rangle$$
satisfies the equation
$$(-\partial^2 +\eta_1^2)G(x)=(1-\frac{\eta_1^2}{\nabla^2})\delta(x).$$
Its Fourier trasform $G(p)$ is thus given by 
$$G(p)=(1+\frac{\eta_1 ^2}{{\bf p^2}})\frac{1}{p^2 +\eta_1 ^2 -i\epsilon}.$$

Of particular interest is the time ordered product
$$G^{\mu}(x)\equiv-\langle 0|(\varphi_1^{\mu}(x)\varphi(0))_+ |0\rangle$$
which is readily seen to satisfy the divergence condition 
$$\partial_{\mu}G^{\mu}(x)=i\delta(x).$$
In a covariant gauge, $G^{\mu}(x)$ must simply be the four-dimensional gradient of a scalar function, and it follows that its Fourier transform is simply
$$G^{\mu}(p) = \frac{p^{\mu} } {p^2 - i \epsilon}.$$
Higgs \cite{Higgs Letters} assumes (erroneously) that local charge conservation implies that in the radiation gauge  there exists a globally conserved charge operator.  This in turn requires the imposition of  unneeded conditions on the form of $G^{\mu}(p)$.    The temporal component of this function in the radiation gauge follows immediately from the relation between $\varphi^0_1$ and $\varphi_1$.  The spatial components obtain from the observation that they  must be the components of a gradient,  which together with the divergence condition on $G^{\mu}$, yields
$$G^{\mu}(p)= \left( p^k (1+\frac{\eta_1^2}{\bf p^2} ), p^0 \right) \frac{1}{p^2 + \eta_1^2 -i\epsilon}.$$
This is readily seen to satisfy the divergence condition,  but this time without the appearance of  a NG boson. 

In sum this shows that massless particles necessarily remain in a covariant formulation of the problem.  These have not (as so frequently misstated) been subsumed into the massive vector meson as its longitudinal component.  These zero mass particles comprise an essential part of the excitation spectrum of the fields that define the model.  That they are in fact harmless gauge excitations follows from the gauge equivalent demonstration in GHK of their absence from the \emph{physical} spectrum. 

\appendix
\section{Appendix}

To facilitate comparison of the above results with those of  Ref.[2] a number of comments are in order concerning the consequences of gauge choice.  One notes that the identification of $\varphi^{\mu}_1$ with $B^\mu$, and $F^{\mu\nu}$ with $G^{\mu\nu}$ yields the set 
$$\partial_\mu B^\mu=0$$
$$\partial_\nu G^{\mu\nu}-\eta_1^2B^\mu=0$$
of Ref.{2}.  While the existence of massive vector mesons  implied by these equations together with the remaining scalar degree of freedom might be thought to comprise the entire content of the action, this is in fact not the case.   As shown explicitly above there \emph{must} exist massless gauge excitations in any manifestly covariant gauge.  For the classical gauge theory explored in Ref.[2] there is the possibility of a massless excitation that depends on the choice of boundary conditions.  On the other hand for a quantum description the commutation relations require a massless excitation if the gauge is chosen to be manifestly covariant.  This excitation is pure gauge and hence does not affect physically measurable results.  This problem does not arise in the radiation gauge, and therefore makes it a compelling choice.  However, this does not eliminate the need to demonstrate the absence of massless gauge particles from the physical sector of the theory when a covariant gauge is employed.

The necessary occurrence of such zero mass excitations in a covariant gauge follows from the fact that the vanishing of the divergence of $B^\mu(x)$ allows additional solutions which differ from any given solution by the addition of a gradient term
$\partial^\mu w(x)$, as long as $\partial^2 w(x)=0$.  While classically $B^\mu(x)$ can be chosen so as to eliminate zero mass excitations, the Goldstone theorem (which comes into play in manifestly covariant theories) allows no easy escape.  Specifically, it requires in the current context that 
$$\langle 0|\int  d{\bf x}[B^0({\bf x}, t), \varphi(0)]|0\rangle=-i,$$
a condition that can only be satisfied for arbitrary $t$ if $B^\mu$ and $\varphi$ have a massless particle in their spectra.  This observation is not made in Ref.[2], and while the zero mass particle is not physical, it must necessarily be present (not "eaten" as stated in [5]).  This result is not new, having been recognized by GHK [3] and the massless gauge excitations have been mentioned in separate works by Kibble and by Weinberg[7]. 

\section{Acknowledgment}

This work was partially supported under DOE grant DE-SC0010010 - Task D.

\end{document}